\documentclass[a4paper,onecolumn,11pt]{quantumarticle}
\pdfoutput=1
\usepackage[utf8]{inputenc}
\usepackage[english]{babel}
\usepackage[T1]{fontenc}
\usepackage{amsmath}
\usepackage{hyperref}

\usepackage{tikz}
\usepackage{lipsum}

\usepackage{fontenc}
\usepackage{newunicodechar}
\newunicodechar{̀}{`}
\usepackage{textcomp}
\usepackage{amsmath}
\usepackage{amssymb}
\usepackage{graphicx}
\usepackage{braket}
\usepackage{url} 
\usepackage{comment} 
\usepackage{subfigure}
\usepackage[makeroom]{cancel}
\usepackage{setspace} 
\usepackage{hyperref}

\usepackage{authblk}
\usepackage[legalpaper, portrait, margin=1in]{geometry}

\usepackage[
    backend=biber,
    style=ieee
]{biblatex}

\begin{document}

\title{Local Many Worlds in Spacetime: Deutsch-Hayden Descriptors and Local Wavefunctions}

\author{Mordecai Waegell}
\affiliation{Institute for Quantum Studies, Chapman University, Orange, CA, USA}
\orcid{0000-0002-1292-6041}
\email{waegell@chapman.edu}

\maketitle

\begin{abstract}
Bell's theorem establishes that a local and spatially separable physical theory with no superdeterminism or retrocausality must be a theory of many local worlds.  The local many worlds formalism of quantum mechanics is a transparent separable physical theory in spacetime with a Lorentz covariant causal structure.  Unlike the Everett many worlds interpretation, there is no nonseparable universal wavefunction in configuration space, but like Everett, there is only local unitary evolution which conserves probability (fluid) current, and no wavefunction collapse.  Each event in spacetime has its own local wavefunction which is built up using only the local wavefunctions and local unitary evolution operators within its past light cone, and which is entirely disconnected from spacelike separated local wavefunctions.   We call this the local Schr\"{o}dinger picture.  Local many worlds can also be described in the Heisenberg picture by ascribing Deutsch-Hayden (DH) descriptors to each individual system, each of which evolves under local unitary operations.  Both the local wavefunctions and DH descriptors contain records of past interactions, and while the local wavefunctions provide a clear and intuitive physical narrative, the descriptors contain all of the same information in a much more compact form.  Unfortunately, the compact form is also somewhat cryptic, which has made the local Heisenberg picture relatively inaccessible.  The purpose of this article is to show that the local wavefunctions for any $N$-qubit system can always be constructed directly from the DH descriptors, allowing us to combine the benefits of both pictures.
\end{abstract}

\section{Introduction}\label{intro}

\pagenumbering{arabic}
\setcounter{page}{1}

In the Local Schr\"{o}dinger (LS) picture \cite{waegell2023local}, the local wavefunctions are ascribed to events in spacetime rather than to specific systems, while local DH descriptors \cite{deutsch2000information, bedard2021cost, Kuypers_2021, bedard2021abc, bedard2025explaining, Kuypers2026Restoring} are ascribed to individual systems, and evolve along those systems worldlines.  The local wavefunction at an event is unique and can be thought of as the image of all local quantum information on the surface of the past light cone, analogous to the instantaneous image seen by any observer in special relativity, which will naturally include an image of all systems where their worldlines intersect the surface of the past light cone.   

In subtle contrast, the DH descriptors of a given system only contain information about other systems in the past interaction cone of that system, but not necessarily every system in the past light cone.  However, if we assume that spacetime is filled with massless particles propagating in every direction, and that these interact at least weakly with other systems when they meet at the same events in spacetime, then the past interaction cone would coincide with the past light cone, and the DH descriptors would include every system in the past light cone.  Regardless, we can always construct the local wavefunction at an event using the DH descriptors of every system at the point where its worldline intersects the surface of that event's backward light cone.  To illustrate this, we proceed with a set of $N$ qubits that fall within the past light cone of a particular spacetime event $(\vec{x},t)$.

The remainder of this letter is organized as follows: Section 2 reviews the LS picture, Section 3 reviews the Local Heinsenberg picture and the DH descriptors, Section 4 shows how wavefunctions can be constructed from DH descriptors, and Section 5 discusses the outlook on future work.

\section{The Local Schr\"{o}dinger Picture}

The local wavefunction at an event $(\vec{x},t)$ is constructed by taking the past light cone $C$ of that event, beginning with the set of local wavefunction on any past boundary of that light cone $C_B$, and then acting all local evolution unitaries $U$ within $C$ on them in temporal order \cite{waegell2023local}.  We also presume the initial local wavefunction of each qubit is $|0\rangle^n$ in order to make the mapping to the Heisenberg picture transparent, but this can be trivially generalized to any other initial state by including the $U$ that maps between them.  We then have the local wavefunction,
\begin{equation}
    |\psi\rangle = \prod_{j\in C} U_{j} \bigotimes_{n \in C_B} |0\rangle^n, \label{LocalWavefunction}
\end{equation}
where the $U_j$ are all of the local unitary evolution operators associated to the events within the past light cone, indexed in temporal order for time/light-like separated events.

The local wavefunctions give a transparently Lorentz covariant quantum theory, where the local many worlds structure of the theory is easy to see, but this construction has an undesirable feature:  In order to correctly build up all of the local wavefunctions, we must track the complete list of local unitaries in $C$, and so this information, in addition to all the the local wavefunctions on the boundary $C_B$ -- and we must make sure that $C_B$ is far enough in the past of any scenario we are considering that all of its local wavefunctions are compatible -- which we will explain later.  All of this quantum information must be carried along in point-like packets at $c$, so that it is always available where it needs to play an explanatory role in the physics.  This is still quantum information, which is not experimentally accessible, so there is no concern about thermodynamic limits, and the story can still be Markovian -- but it is nevertheless a cumbersome description.

Quantum information spreads at the speeds of light in the LS picture (as opposed to spreading instantaneously in standard quantum theory), but this should not be confused with decoherence, which is a matter of how systems are entangled: As long as two systems are not entangled in the local wavefunction, they do not appear to have decohered with respect to each other.

The LS picture belongs to a body of work on local quantum theory that originated with the Parallel Lives framework of Brassard and Raymond-Robichaud \cite{ brassard2012can, waegell2017locally, waegell2018ontology, brassard2019parallel} and has grown to include a local fluid treatment in spacetime, inspired by Madelung \cite{madelung1927quantum, waegell2024toward, waegell2024madelung}, which is related to Schr\"{o}dinger's original attempt to get a matter density field in spacetime from his new configuration space theory \cite{Schrodinger1926, schrodinger1927abhandlungen, Allori2010ManyWorlds}.  It is also logically equivalent to the Lorentz covariant quantum field theory developed independently by Schwinger \cite{Schwinger1948Quantum} and Tomonaga \cite{Tomonaga1946}, although formal mathematical equivalence has yet to be shown.

\section{The Local Heisenberg Picture}
Locality is imposed in this picture by the assumption that spacelike separated observables must commute.  Our $N$ qubits have $N$ separate Deutsch-Hayden descriptors \cite{deutsch2000information, bedard2021abc}, each of the form $\mathbf{q^n} = (A,B)$, with $n=1,\ldots,N$, where the first operator $A$ is understood as the $\sigma_x$-component $q_x$ of the descriptor, the second operator $B$ is understood as the $\sigma_z$-component $q_z$, and it is implied that the $\sigma_y$-component $q_y$ is $iAB$.  By construction, the DH descriptors of different systems are mutually commuting, even when those systems are entangled. A system's descriptor is a localized point-like packet of quantum information that the system carries along its worldline, and it evolves only when local unitary operations (including interactions) act on the system, resulting in a Lorentz covariant causal structure.  When multiple systems locally interact, their descriptors generally mix and become operators on the joint Hilbert space of all systems, creating a quantum information record of that interaction between those systems.  Descriptors with Hilbert spaces for multiple qubits are generally entangled, and when entangled systems interact, the descriptors produce expectation values (using the fixed Heisenberg state $|0\rangle^{\otimes N}$) that obey the entanglement correlations.

The DH descriptors contain all the information needed to construct the local wavefunctions, but have the advantage that they need not carry around the full history of interaction unitaries in $C$, nor the initial descriptors on $C_B$ -- one simply acts the local unitaries upon the DH descriptors as they come, and this encodes all of the necessary information.  Thus, provided we take the descriptors of each system at the point where it intersects the surface of the past light cone $C_S$, they will have already encoded the effects of every $U_j$ in $C$ on the initial descriptor values.

We can use the same information to construct the local reduced density matrix for any subset of the $N$ qubits using only the descriptors of the systems in the subset (although one must generally use the Heisenberg states $|0\rangle^n$ for every qubit that belongs those descriptors' Hilbert spaces). 

It is worth noting that the spatially separable DH descriptors can also be used to calculate a standard quantum wavefunction on a spacelike hyperplane by taking the descriptors where those systems' worldlines intersect that hyperplane.  This helps to highlight the nonlocality of the standard nonseparable wavefunction in configuration space: in order to construct it from Lorentz covariant components, we need to explicitly collect pieces of quantum information from many spacelike separated points.  This also shows why these wavefunctions are not Lorentz covariant: the standard wavefunction on a different spacelike hyperplane will be built from descriptor values at different events.  Furthermore, doing this creates the illusion that entangled systems are actively connected at spacelike separation, rather than simply sharing some interaction history.  By contrast, when one constructs local wavefunctions using the descriptor values on $C_S$, separable locality and Lorentz covariance are preserved.

\section{Constructing Wavefunctions from DH Descriptors}

Any $N$-qubit density matrix can be represented as
\begin{equation}
    \rho = \frac{1}{2^N}\bigg(I + \sum_{j=1}^{4^N-1} c_j P_j   \bigg),\label{Rho}
\end{equation}
where the $P_j$ are the set of $N$-qubit Pauli operators constructed as tensor products of Pauli matrices with each other and/or the $2\times 2$ identity, which is a spanning set (indexed in no particular order - for example, it could be that for 3 qubits, $P_1 = \sigma^1_x \otimes \sigma^2_x \otimes \sigma^3_x$.),  and the
\begin{equation}
    c_j = \textrm{Tr}\big(\rho P_j\big)
\end{equation}
are real coefficients.  If the state is pure, i.e., $\rho = |\psi\rangle \langle \psi|$, we also have $\sum c_j^2 = 2^N-1$, and we can get $|\psi\rangle$ by taking any nonzero column of $\rho$ and renormalizing, i.e., $\rho|m\rangle = |\psi\rangle \langle \psi|m\rangle$, and thus  provided $\langle\psi|m\rangle \neq 0$, we have $|\psi\rangle =\rho|m\rangle/\langle \psi|m\rangle$.

We construct the local state $\rho$ at $(\vec{x},t)$ using the descriptors $\mathbf{q}^n$ of each of the $N$ qubits where its worldline intersect $C_S$.  We next construct the descriptor analog $Q_j$ of each $N$-qubit Pauli matrix, e.g., if $P_7 = \sigma^1_z \otimes \sigma^2_x \otimes \sigma^3_z$, then $Q_7 = q_z^1 q_x^2 q_z^3$ (noting again that the descriptor components of different qubits mutually commute, even if they contain operators for the same qubits, so the order of this product is arbitrary), which gives $c_j = \langle 0 |^{\otimes N} Q_j |0 \rangle^{\otimes N}$.  We have thus constructed $\rho$ and the local wavefunction $|\psi\rangle$ from the DH descriptors.  

For example, if the initial local wavefunction of a qubit is $|0\rangle$ (the same as the the Heisenberg state), then the initial descriptor is $\mathbf{q} = (X,Z)$, where $X$ and $Z$ are the corresponding Pauli matrices, and $y$-component is $Y = iXZ$.  Applying the Heisenberg state to these descriptors gives us $c_X = c_Y = 0$ and $c_Z = 1$, so $\rho = \frac{1}{2}(I + Z) = |0\rangle\langle 0|$, which returns the initial state, as expected.  A different original state results in different initial descriptors, leaving the Heisenberg state fixed.

This means that we can replace the undesirably cumbersome records of unitaries in $C$ and local wavefunctions on $C_B$ of the LS picture with the DH descriptors, which are much more compact.  In turn, this means that the somewhat cryptic form of the DH descriptors in the local Heisenberg picture automatically gives rise to an intuitive Lorentz-covariant narrative of local wavefunctions evolving in spacetime.  Thus, it seems that the ideal local formulation is a hybrid Schr\"{o}dinger-Heisenberg picture.

These results show that the local information in the DH descriptors is a compression of the local information in the LS picture - in the latter, one must keep track of all $U_j$ within the past light cone in order to construct $|\psi\rangle$, which is the complete interaction history of all $N$ qubits, whereas in the DH picture, one only needs the current descriptors of each of the $N$ qubits at the moment where its worldline intersects $C_S$.  Another way to see this is that given all of of the initial states and local unitaries in the local Schr\"{o}dinger picture, one can always construct the DH descriptors where the intersect $C_S$, but given those descriptors, one cannot reconstruct the full list of $U_j$, since there is no unique unitary history that would lead to those descriptors.

\section{A Bell Experiment}
\begin{figure}[ht!]
    \centering
    \includegraphics[width=\linewidth]{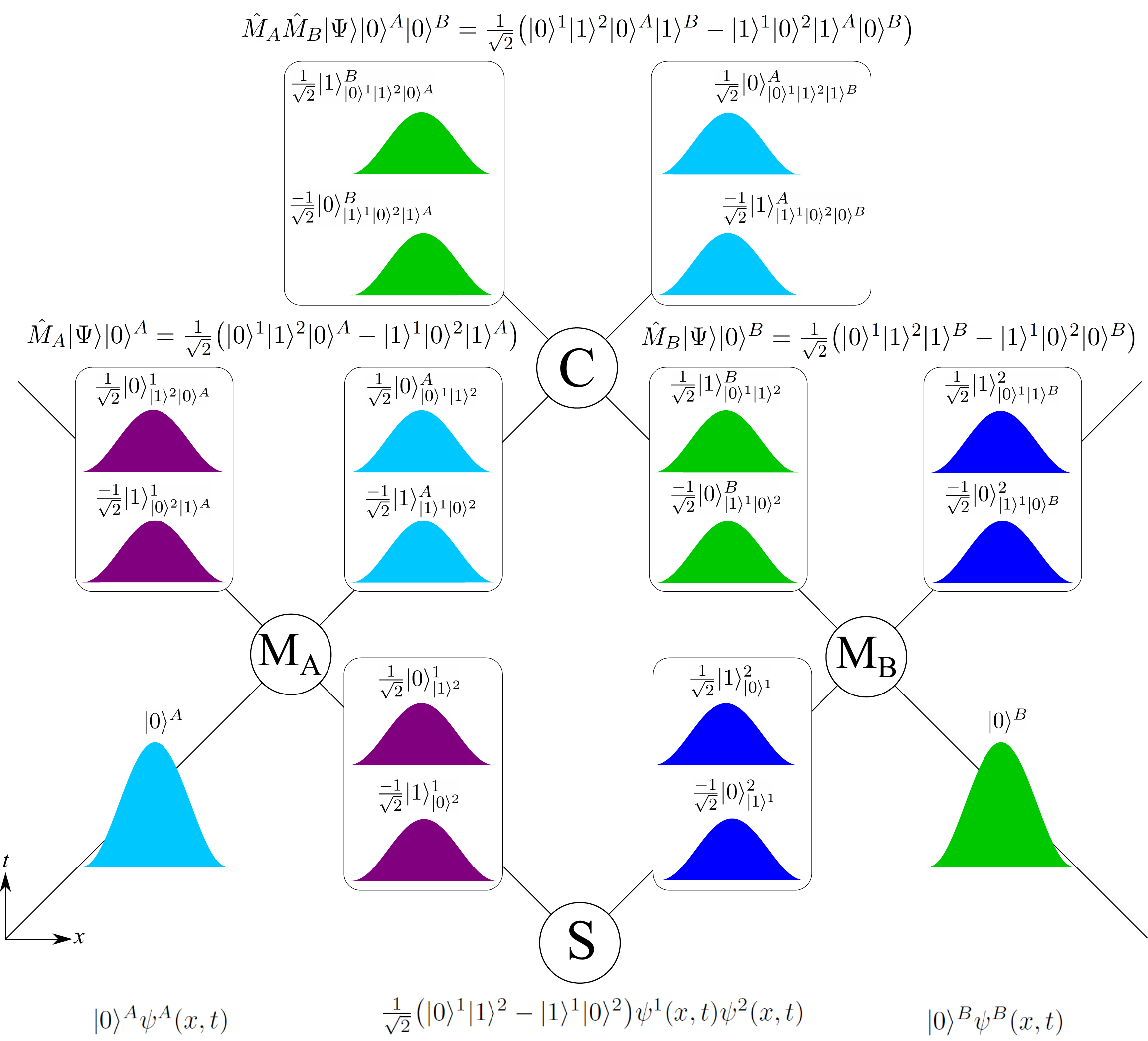}
    \caption{Spacetime diagram of a Bell experiment for a run where Alice and Bob measure the same direction.  Each system is represented by a conserved fluid which moves on a straight trajectory through the experiment.  These wave packets do not become entangled with the qubits they carry during this experiment.  At the beginning of the experiment, the descriptors of Alice, Bob, and the qubits in the singlet state are  $\mathbf{q}_A = ( X^A, Z^A)$,  $\mathbf{q}_B = (X^B, Z^B)$, $\mathbf{q}_1 = (-Z^1 X^2, -X^1 I^2)$, $\mathbf{q}_2 = (I^1 X^2, -X^1 Z^2)$, respectively, where $X^n$, $Y^n$ and $Z^n$ are the Pauli matrices, and $I^n$ the identity for qubit $n$.  It is easy to see that these give the local wavefunctions shown.  When Alice measures qubit 1 using $\hat{M}_A = \textrm{CNOT}_{1 \rightarrow A}$, their descriptors are updated to $\mathbf{q}'_A = ( I^1 X^A, X^1 Z^A)$ and $\mathbf{q}'_1 = (-Z^1 X^2 X^A, -X^1 I^2 I^A)$, but the descriptors of Bob and qubit 2 are entirely unaffected.  If we consider Alice's past light cone $C$ just after her measurement, and take the descriptors $\mathbf{q}'_A$, $\mathbf{q}'_1$, and $\mathbf{q}_2$ where the respective worldlines of Alice and qubits 1 and 2 cross $C_S$, we can construct the local wavefunction $\frac{1}{\sqrt{2}}\big(|0\rangle^1|1\rangle^2|0\rangle^A - |1\rangle^1|0\rangle^2|1\rangle^A\big)$.  Likewise, when Bob measures qubit 2 using $\hat{M}_B = \textrm{CNOT}_{2 \rightarrow B}$,  their descriptors are updated to  $\mathbf{q}'_B = (I^2 X^B, -X^1Z^2 Z^B)$ and $\mathbf{q}'_2 = (I^1 X^2 X^B, -X^1 Z^2 I^B)$, and the descriptors of Alice and qubit 1 are entirely unaffected.   If we consider Bob's past light cone $C$ just after his measurement, and take the descriptors $\mathbf{q}'_B$, $\mathbf{q}'_2$, and $\mathbf{q}_1$ where the respective worldlines of Bob and qubits 1 and 2 cross $C_S$, we can construct the local wavefunction $\frac{1}{\sqrt{2}}\big(|0\rangle^1|1\rangle^2|1\rangle^B - |1\rangle^1|0\rangle^2|0\rangle^B\big)$.  Finally if consider either Alice's or Bob's past light cone $C$ after they have met to compare results at event C, we can use the descriptors of all four systems where they cross the $C_S$ to construct $\frac{1}{\sqrt{2}}\big(|0\rangle^1|1\rangle^2|0\rangle^A|1\rangle^B - |1\rangle^1|0\rangle^2|1\rangle^A|0\rangle^B\big)$.} 
    \label{fig:Same}
\end{figure}

\begin{figure}[hb!]
    \centering
    \includegraphics[width=\linewidth]{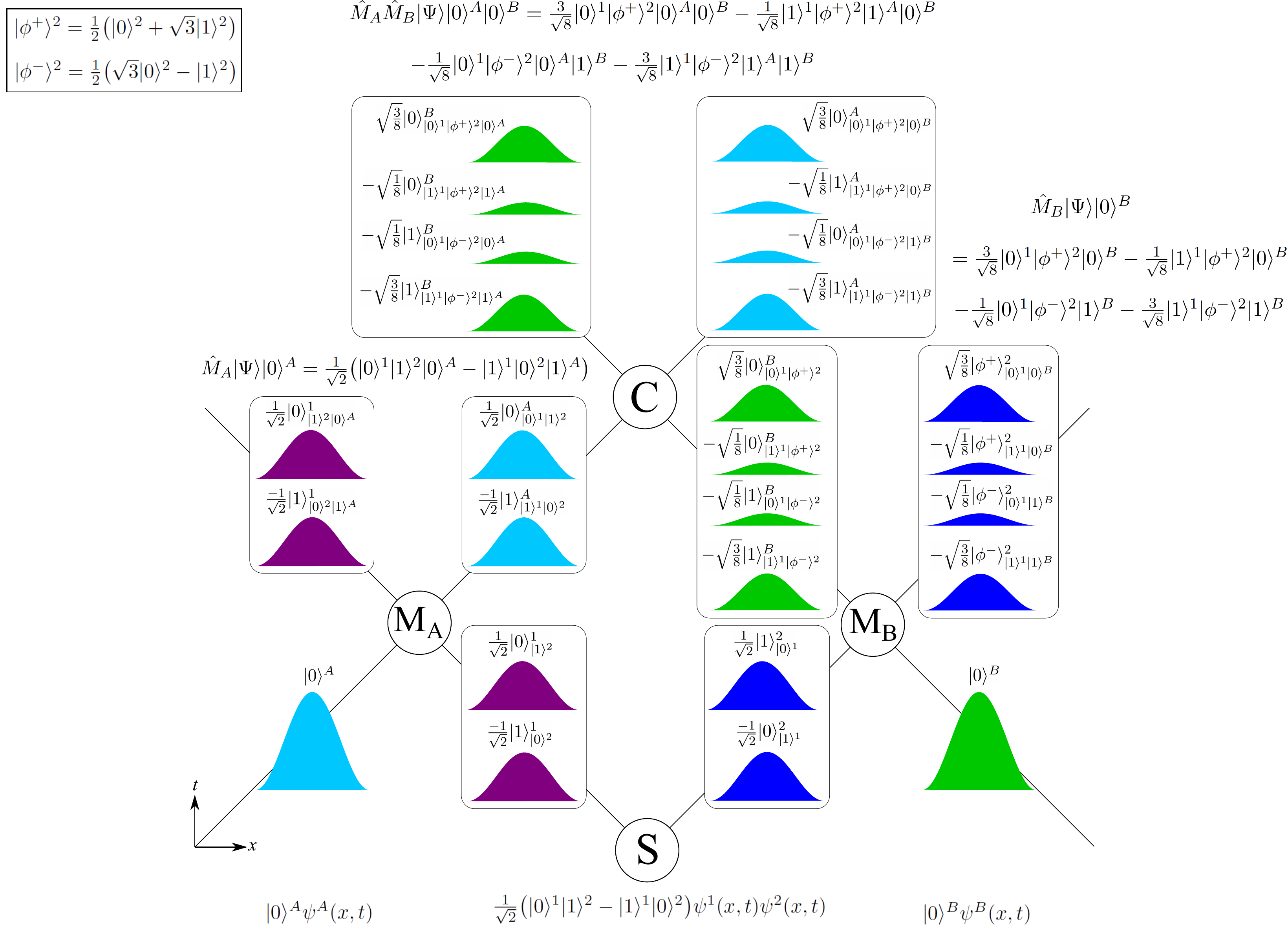}
    \caption{Spacetime diagram of a Bell experiment for a run where Alice and Bob measure directions 120$^\circ$ apart in the $xz$-plane.  The descriptors at the beginning and after Alice's measurement are the same as in Fig. \ref{fig:Same}, but after Bob's measurement,his descriptor is $\mathbf{q}'_B = \big(I^1 I^2 X^B, \frac{\sqrt{3}}{2}I^1 X^2 Z^B + \frac{1}{2}X^1Z^2Z^B  \big)$, and the descriptor of qubit 2 is $\mathbf{q}'_2 = \frac{1}{4} \big(I^1 X^2 X^B - \sqrt{3}X^1Z^2X^B + 3X^1X^2I^B + \sqrt{3}X^1Z^2I^B    ,   \sqrt{3} I^1 X^2 X^B - 3X^1Z^2X^B - \sqrt{3}X^1X^2I^B - X^1Z^2I^B  \big)$.  The measurement unitary used for this update decomposes as $\hat{M}_B = R^2_y\cdot(\frac{2\pi}{3}) \cdot Z^2 \cdot CNOT_{2\rightarrow B} \cdot R^2_y(\frac{2\pi}{3}
    )\cdot Z^2$.  The descriptors may look cumbersome in this form, but each component always reduces to a single operator on the joint Hilbert space.  If we consider Bob's past light cone $C$ just after her measurement, and take the descriptors $\mathbf{q}'_B$, $\mathbf{q}'_2$, and $\mathbf{q}_1$ where the respective worldlines of Bob and qubits 1 and 2 cross $C_S$, we can construct the local wavefunction $\frac{1}{\sqrt{8}}\big(3|0\rangle^1|\phi^+\rangle^2|0\rangle^B - |1\rangle^1|\phi^+\rangle^2|0\rangle^B  - |0\rangle^1|\phi^-\rangle^2|1\rangle^B  - 3|1\rangle^1|\phi^-\rangle^2|1\rangle^B\big)$.  Finally if consider either Alice's or Bob's past light cone $C$ after they have met to compare results at event C, we can use the descriptors of all four systems where they cross $C_S$ to construct $\frac{1}{\sqrt{8}}\big(3|0\rangle^1|\phi^+\rangle^2|0\rangle^A|0\rangle^B - |1\rangle^1|\phi^+\rangle^2|1\rangle^A|0\rangle^B  - |0\rangle^1|\phi^-\rangle^2|0\rangle^A|1\rangle^B  - 3|1\rangle^1|\phi^-\rangle^2|1\rangle^A|1\rangle^B\big)$.  
    }
    \label{fig:Diff}
\end{figure}

To really see the Lorentz covariant ontology of the local many worlds theory at work, it is best to go through a Bell experiment \cite{bell1964einstein, Bell2} to see how the empirical entanglement correlations are produced using DH descriptors (see also \cite{waegell2023local, bedard2025explaining}).  As an illustrative example, we use the 2-qubit Bell test popularized by Wigner and Mermin \cite{wigner1970hidden, mermin1981bringing, mermin1985moon}, wherein a singlet state is prepared, and then one qubit is sent to Alice and the other to Bob, in spacelike separated regions, whereupon each of them freely/randomly chooses to measure their qubit's spin in one of the same three equally spaced directions in the $xz$-plane.  If one assumes no superdeterminism or retrocausality is at work here, and the other tacit assumptions of Bell's theorem, then it follows that the empirical statistics of this experiment cannot have a locally causal explanation, and thus they violate a Bell inequality.

One of the tacit assumptions of Bell's theorem is that there is only one world, so that there must exist a single Alice and a single Bob who have observed correlated results, even while they are spacelike separated -- which then requires the existence of the usual single-valued local hidden variables, which leads to a logical contradiction.   Note that having many global worlds, does not improve this situation, since the Alice and Bob in each world must still be correlated at spacelike separation \cite{WAEGELL2020} - and the same basic problem extends to any quantum theory with a spatially nonseparable density matrix \cite{Waegell2026Nonlocal}, as in Everett many worlds \cite{everett1957relative}, or Spacetime State Realism \cite{wallace2010quantum}.  However, if there are only separable local worlds, defined along many worldlines in one spacetime, then we can have multiple Alices in her region, having seen all possible outcomes, and likewise multiple Bobs in his region, and matching them up according to the entanglement correlations can wait until they locally interact to compare results.  Thus, locally mediated quantum information can explain all of the observed outcomes for all of the Alices and Bobs in this experiment, including the entanglement correlations.  This is why modern versions of Bell's theorem require an explicit `one world' assumption \cite{WAEGELL2020}, and thus prove that a local theory without superdeterminism or retrocausality must have many local worlds of the type described here.

We approximate Alice and Bob as qubits in this example, since their microscopic details do not play a significant role.  Because of the symmetry of this experiment, we only need to consider two different scenarios in order to demonstrate how we can get all of the measurement statistics in local many worlds using locally mediated quantum information; the case where Alice and Bob measure the same direction , and the case where they measure directions that are 120$^\circ$ apart in the $xz$-plane, shown in Fig. \ref{fig:Same} and Fig. \ref{fig:Diff}, respectively.

There figures show the initial local wavefunctions at the bottom, and then as time progresses upward, the local wavefunctions after the local measurement unitaries are shown in the middle, and finally the local wavefunction after Alice and Bob meet to compare their results at the top.  The DH descriptors are are given in the figure captions, and make use of the following unitary update rules \cite{bedard2021abc}:
\begin{equation}
    \begin{array}{ccccc}
       (A,B) \xrightarrow{Z} (-A,B),  && (A,B) \xrightarrow{X} (A,-B), && (A,B) \xrightarrow{H} (B,A),
    \end{array}
\end{equation}
\begin{equation}
    \begin{array}{c}
(A,B) \xrightarrow{R_y(\theta)} (A\cos\theta + B\sin\theta, -A\sin\theta + B\cos\theta), 
    \end{array}\nonumber
\end{equation}
\begin{equation}
    \left\{\begin{array}{c}
         (A^c,B^c)  \\
         (A^t,B^t) 
    \end{array}\right\} \xrightarrow{\textrm{CNOT}}
       \left\{\begin{array}{c}
         (A^c A^t,B^c)  \\
         (A^t,B^c B^t) 
    \end{array}\right\},\nonumber
\end{equation}
where $R_y(\theta)$ is the rotation operator for angle $\theta$ about the $+y$-axis, and $H$ is a Hadamard,  and $c$ and $t$ denote the control and target qubits of the CNOT, respectively.  The initial singlet descriptors are obtained by applying unitary $Z^1\cdot\textrm{CNOT}_{1\rightarrow 2}\cdot H^1\cdot X^2$ to the Heisenberg states of qubits 1 and 2.

The position basis is fundamentally preferred in local many worlds, so the macroscopic pointer needle points to a single spot in each world.  The different worlds are shown for Alice and Bob, where the normal script describes the observer's distinct empirical experience in that local world (also called the external memory), and the subscripts describe records of past interactions.  Note that promoting all subscripts to normal script and summing the different worlds of a given system reproduces the local wavefunction (also called internal memory).  The conserved probability current for each system is reinterpreted in this model as the current for a fluid of many local worlds, each a point-like particle on a worldline in spacetime.  The amount of fluid in each local world is given by the modulus square of its coefficient.  The fluid is conserved and always divides with Born rule proportions, which gives rise to empirical frequencies that approximate the Born rule for typical observers \cite{Waegell2025From, Waegell2026OnMeasurement}. 

When the many local copies of Alice and Bob meet at event C, the local wavefunction there determines what proportion of Alice's with each outcome meet which proportions of Bob's outcomes, such that the observations of all Alices and Bobs obey the proper entanglement correlation.  As discussed above, this locally causal delayed matching mechanism for producing the entanglement correlations requires the existence of multiple Alices and Bobs at spacelike separation.

Another interesting way to see the mechanism of delayed matching is as an enforcer of conservation laws.  After she measures, there are Alices on her side who saw up, and Alices who saw down, and likewise for the Bobs on his side.  Consider a run where Alice and Bob measure the spin in the same direction.  The singlet state has spin zero, so if an Alice who saw spin up later meets a Bob who also saw spin up, then conservation of angular momentum would be violated -- but this is prevented because the local wavefunction produces delayed matching which obeys the entanglement correlations.  This connection between entanglement correlations and conservation laws appears to be quite general.

Finally, this example also allows us to illustrate the issue of incompatible versus compatible local wavefunctions in the LS picture.  Note that in Figs. \ref{fig:Same} and \ref{fig:Diff}, the (expanded) local wavefunction after Alice and Bob have measured are entangled with qubits 1 and 2 in incompatible ways (e.g., in Fig. \ref{fig:Same}, $\frac{1}{\sqrt{2}}\big(|0\rangle^1|1\rangle^2|0\rangle^A - |1\rangle^1|0\rangle^2|1\rangle^A\big)$ on Alice's side, and  $\frac{1}{\sqrt{2}}\big(|0\rangle^1|1\rangle^2|1\rangle^B - |1\rangle^1|0\rangle^2|0\rangle^B\big)$ on Bob's), meaning that when Alice and Bob meet, there is no reliable way to combine those two states to get the correct local wavefunction  (e.g., in Fig. \ref{fig:Same}, $\frac{1}{\sqrt{2}}\big(|0\rangle^1|1\rangle^2|0\rangle^A|1\rangle^B - |1\rangle^1|0\rangle^2|1\rangle^A|0\rangle^B\big)$), unless one knows the unitaries $\hat{M}_A$ and $\hat{M}_B$.  This has two undesirable consequences for the LS picture, as discussed above.  The first is that we need to we need local information packets that carry the full list of local unitaries in $C$ and local wavefunctions on $C_B$, and the second is that we cannot choose just any boundary $C_B$ in order to get the local wavefunction from Eq.\ref{LocalWavefunction}, i.e., we cannot choose $C_B$ with incompatible local wavefunctions.  The wavefunctions are compatible at the bottom of Figs \ref{fig:Same} and \ref{fig:Diff}, but that is not longer the case after Alice's or Bob's measurements, and we cannot get the final local wavefunction at the top of the figures choosing a $C_B$ with those local wavefunctions.  Both of these issues are entirely avoided when we construct the local wavefunctions from the DH descriptors, and they still provide the same transparent many worlds narrative.

\section{Outlook}

The DH descriptors provide a local Lorentz covariant quantum theory where all of the physics occurs in spacetime, but there is still a lot of work to be done fleshing this out, and rehashing the historical development of modern quantum physics with this new perspective.  The local wavefunctions provide an intuitive narrative story about many local worlds, and the Madelung fluid picture should guide us in developing the treatment for spatial degrees of freedom like position and momentum, rather than just qubits.  The particle dynamics in the Madelung equations are classical, but there is already some work on relativistic generalizations \cite{poirier2012trajectory,poirier2020trajectory, ruiz2021direct, sato2024quantum,  reddiger2024towards,   fabbri2025madelung, poirier2025relativistic, sato2024quantum}.   The present approach should work well for fermions, but it will need to be generalized to treat bosons and wavefunctions with indefinite particle number, and we may need to use Fock spaces to get a properly relativistic local fluid picture (see Sec 6.1 of \cite{waegell2023local} on the locality of collapse).  

One major hope for this program is that a local Lorentz covariant quantum theory based on point particles in spacetime should be could lead to a viable quantum gravity theory, and that new progress can be made by starting over at the beginning of quantum physics and using DH descriptors to develop a coherent physical narrative that is consistent with relativity from the outset (see \cite{rubin2022quantum}).  Following Einstein \cite{Einstein1935EPR}, we insist that physics tell us a comprehensible local story about events in spacetime, and as we have shown, local many worlds provides this -- without any appeal to superdeterminism, retrocausality, or any other loopholes in Bell's theorem.  \newline

\noindent \textbf{Acknowledgments:}  This project/publication was made possible through the support of Grant 63209 from the John Templeton Foundation. The opinions expressed in this publication are those of the authors and do not necessarily reflect the views of the John Templeton Foundation.\newline

\noindent\textbf{Conflicts of Interest:} The corresponding author states that there is no conflict of interest. \newline

\noindent\textbf{Data Availability:} There is no data associated with this manuscript.

\printbibliography
\end{document}